\documentclass[9pt,twocolumn,preprint]{article}

\usepackage{hyperref}
\usepackage{lineno}
\usepackage{orcidlink}
\usepackage{authblk}

\usepackage{stfloats}
\usepackage{biblatex}         
\AtEveryBibitem{\clearfield{title}} 

\title{Coherent two-dimensional electronic spectroscopy integrated with confocal back focal plane microscopy}

\author[1,**]{Trideep Kawde\orcidlink{0009-0000-4113-1080}}
\author[1,**]{Pavel Trofimov\orcidlink{0000-0002-6426-2670}}
\author[1]{Anton Trenczek
\orcidlink{0009-0003-0790-0669}}
\author[1]{Matteo Russo \orcidlink{0009-0006-4596-7380}}
\author[1]{Jasper Wilhelm Schwering \orcidlink{0009-0006-4873-2464}}
\author[1,*]{H\'el\`ene Seiler  \orcidlink{0000-0003-1521-4418}}

\affil[1]{Physics department, Freie Universität Berlin, Arnimallee 14, 14195 Berlin}

\affil[*]{helene.seiler@fu-berlin.de}

\affil[**]{Contributed equally}

\begin{document}
\maketitle

\begin{abstract}
\textbf{We introduce a setup for coherent two-dimensional electronic spectroscopy in the pump-probe reflection geometry that is integrated with a confocal back focal plane imaging microscope. The angle-resolved capability is utilized to control pump and probe wavevectors, while real space imaging enables co-localization of the collection spots for linear and ultrafast experiments. Compression of pulses down to 20 fs is achieved. We demonstrate the capabilities of this approach on an exfoliated WSe$_2$ monolayer on Si/SiO$_2$. The setup is suited to investigate excitons and exciton-polaritons in 2D Materials and their heterostructures.}
\end{abstract}


\section{Introduction}
Coherent two-dimensional electronic spectroscopy (2DES) is a powerful tool to reveal couplings between quantum states and disentangle homogeneous and inhomogeneous contributions to spectroscopic lineshapes \cite{zanni}. The rise of 2D materials and their combination in heterostructures provides unique opportunities for the method \cite{Novoselov2016}. Over the last decade, 2DES has become a popular tool to reveal excitonic processes in transition metal dichalcogenide monolayers \cite{Guo2018, Hao2016, Moody2015, Lloyd2021, Timmer2024} and charge transfer processes in heterostructures \cite{Policht2021, Purz2021, Purz2022}. Building on these advances, 2DES holds substantial untapped potential for probing more complex systems, such as twisted bilayers, gated van der Waals devices, or 2D materials integrated with metasurfaces \cite{Li2022, Timmer2026}. A common challenge of working with such samples, in particular exfoliated materials, is their typical sizes of 10-20 $\mu m$. This adds another level of complexity to the experiment since conventionally 2DES is conducted without high-magnification optics. Most previous studies on 2D materials were performed on samples grown by chemical vapor deposition \cite{Moody2015, Guo2018, Hao2016, Lloyd2021, Policht2021} or on large exfoliated flakes of 50 $\mu m$ or more. Moreover, samples are often produced on opaque substrates in the visible range, such as Si/SiO$_2$, thereby preventing the use of the transmission geometry most commonly used in 2DES. Therefore, it is desirable to develop methods that can overcome these challenges.

Coupling confocal or widefield optical microscopes with ultrafast experiments is a natural way to increase spatial resolution in ultrafast optical experiments. Transient-absorption microscopies have been carried out in a variety of ways, and we refer the reader to recent review \cite{Gross2023}. Extending the techniques to 2D spectroscopy is more challenging due to the addition of pulses and dimensions to scan. Few realizations of widefield 2DES microscopy in the pump-probe geometry have been demonstrated by placing the microscope objective after the sample \cite{Azarm2026_arxiv, Steves2019}, making this approach well suited for spatially-resolved 2DES in a transmission geometry. Most implementations of 2D microscopy operate in a fully collinear geometry \cite{Baiz2014, Goetz2018, Tiwari2018, Ostrander2016, Jones2019, Purz2022}. By filling the entire numerical aperture of the objective, these setups optimize for spatial resolution. However, isolating the 2D signals typically involves numerous phase-cycling steps \cite{Baiz2014, Goetz2018, Ostrander2016} or relies on color/polarization filtering \cite{Jones2019}. Moreover, optimizing for the highest spatial resolution comes at the cost of losing angular resolution, making it difficult to disentangle the angle-dependent response of the system. These recent approaches motivate further exploration of the trade-offs between spatial and angular resolution, experimental complexity and flexibility.

Here we introduce a broadband visible 2DES setup in the pump-probe reflection geometry that is integrated with a confocal Fourier optical microscope. This microscope, based on a high numerical aperture apochromat objective, is used to perform imaging and various linear spectroscopies at the same position as the 2DES measurements. Real and back focal plane imaging can be utilized for filtering of the 2D signals and quantitative determination of pump and probe wavevectors. We demonstrate the capabilities of our instrument on a prototypical 2D material, an exfoliated monolayer of WSe$_2$ on Si/SiO$_2$. Our setup opens up studies on 2D materials and heterostructures with sizes down to 10-20 $\mu m$, while retaining full flexibility in pump and probe color, polarization as well as experiment geometry (transmission, reflection). We envision its back focal plane imaging capabilities to be particularly relevant for future ultrafast studies on polaritonic systems \cite{Xu2023}.

\section{Experimental setup}

\begin{figure*}[t]
\centering
\includegraphics[width=1\linewidth]{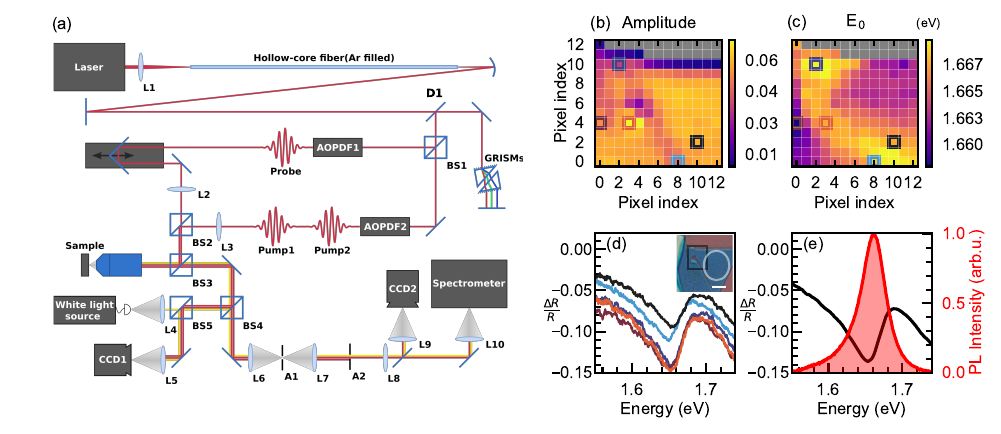}
\vspace{-10pt} 
\caption{ Instrument overview and linear properties of the sample. (a) Sketch of the setup combining 2DES with a back focal plane imaging microscope. Linear reflectivity maps showing (b) exciton amplitude and (c) exciton energy E$_0$ measured with a spatial resolution of 1 $\mu m$. (d) Exemplary local linear reflectivity spectra, color-coded by small squares in (b) and (c). Inset: Real Image of the sample (scale bar = 10 $\mu m$). The black square shows the region where reflectivity mapping is performed. (e) Linear reflectivity (black) and photoluminescence (red) spectra acquired at the same position as 2DES measurements (white circle in inset (d)). }

\label{fig1}
\end{figure*}

A sketch introducing the key components of the experiment is shown in Fig. \ref{fig1}(a). The optical setup is designed to conduct 2DES while adding the functionality of confocal Fourier optical microscopy at the same sample position. 
The output pulses from a Ti:sapphire regenerative amplifier (Astrella, Coherent, 1 kHz, 70 fs) are attenuated and sent to a 2.35 m long hollow-core fiber setup (Few-cycle) filled with argon to a pressure of 2 bar for spectral broadening, similar to previous reports \cite{Seiler2017, Sonnichsen2021, Timmer2023, Ma2016, AlHaddad2015}. Alternatively, the fiber can be pumped with the output of a commercial optical parametric amplifier (TOPAS, 0.47 eV - 2.61 eV). The input beam to the fiber is actively stabilized (Aligna, TEM Messtechnik).

The spectrally broadened pulses generated in the fiber are collimated and split into pump and probe channels via a beamsplitter (BS1), each containing an acousto-optic programmable pulse shaper (WR25 Dazzler, Fastlite, working range: 1.30 -2.69 eV) \cite{Verluise2000}, AOPDF1 and AOPDF2 in Fig. \ref{fig1}(a). As an option, a pair of GRating and PrISMs (GRISMs) can be used for dispersion pre-compensation of the pulses when working with energies higher than 1.77 eV. The laser pulses are coupled to the GRISMs using a horizontal D-shaped mirror (D1). The pulse shaper in the pump arm is used to generate two phase-locked collinear pump pulses with well-defined coherence time (t$_1$) and relative carrier envelope phase (CEP). Both shapers are also employed for pre-compensation of the significant chirp generated in the thick apochromat glassy objective \cite{Pawowska2014}. To scan through population time (t$_2$), probe pulses are delayed using a motorized 3 ns delay stage (Newport, DL225).

While the setup features both a transmission and a reflection channel, we decide to focus here on the less common reflection geometry. Pump and probe pulses are combined in the beamsplitter (BS2) in the chosen operating channel and coupled to a high numerical aperture apochromat objective (Mitutoyo 50x, NA 0.42, working distance 17 mm), providing a broad range of excitation angles. Lenses L2 and L3 are used to control the incident angles and beam sizes of pump and probe beams in the back focal plane of the objective.  The pump and probe pulses are characterized at the sample position using a combination of chirp scans and a home-built transient-grating frequency-resolved optical gating (tg-FROG) setup \cite{Loriot2013, Trebino1997}. Specifically, for pulse characterization, the pump and probe beams behind the sample position are collimated by a 0.5" diameter silver off-axis parabolic mirror with an effective numerical aperture (NA) of 0.44. The beam is then routed towards the characterization setups. To facilitate the search for optimal compression parameters, chirp scans are performed first to find the optimal second-order coefficient to be applied by the pulse shapers \cite{Loriot2013}. Residual chirp as well as third and fourth order spectral phase distortions are subsequently analyzed with the tg-FROG setup and further corrected using the combination of AOPDFs and GRISMs.

The generated 2D signals from the sample are collected with the same apochromat objective and coupled through beamsplitter (BS2) towards a custom built infinity corrected back focal plane confocal optical microscope. The signals can be spatially filtered in the intermediate real image of the sample (A1). Alternatively, an aperture placed in the intermediate Fourier image (A2) can be employed to block the pump or increase angular resolution (Fig. S1). Adding/Removing L8 enables imaging either the real/back focal plane into the detector and CCD2.  
In the pump-probe geometry, the desired rephasing and non-rephasing 2D signals are emitted alongside the undesired pump-probe and linear signals in the probe direction \cite{Myers2008, Shim2009}. To eliminate those contributions, a two-step phase-cycling procedure is carried out. The nonlinear reflectivity is obtained as ((R$_{ON}$($\phi_{12} = \pi$) - R$_{ON}$($\phi_{12} = 0$))/R$_{OFF}$, where R$_{ON}$ and R$_{OFF}$ are reflectivity with and without pumps and $\phi_{12}$ is the CEP difference between pulse 1 and 2. This signal, homodyne detected with the third pulse, is imaged (L10) towards the entrance slit of the detection spectrometer (Stresing, Hamamatsu S12600-1006 CCD sensor), thereby performing the Fourier transform along the detection time axis.

In addition to the 2DES capabilities, the detection part of the setup is designed to conduct imaging and linear spectroscopies (transmission, reflectivity, and photoluminescence) of the sample at the same position as 2DES measurements are performed \cite{Steves2019}. This aspect of the spectrometer makes it particularly well-suited to investigate 2D materials, in view of their high heterogeneity. Sample mapping is performed using automated raster scans with piezoelectric stages, enabling direct visualization of the spatially‑distributed spectral inhomogeneities. A continuous wave light source (laser or stabilized white light) and imaging cameras (CCD1 for real space imaging and CCD2 with L8 for back focal plane imaging) are optically coupled using beamsplitters (BS4, BS5) to the main part of the optical microscope. 

\section{Example: monolayer WSe$_2$}

We demonstrate the capabilities of our setup using a mechanically exfoliated monolayer of WSe$_2$ on Si/SiO$_2$, a prototypical 2D material. A real space image of the sample is shown in the inset of Fig. \ref{fig1}(d), and the monolayer is identified by AFM measurements as the blue triangle in the image. Before the ultrafast experiments, linear reflectivity of the monolayer is measured at the sample position (black curve in Fig. \ref{fig1}(e)). The spectra feature a characteristic peak of the A exciton of WSe$_2$ around 1.667 eV. A photoluminescence spectrum acquired at the same position is overlaid (shaded red). Furthermore, mapping linear reflectivity on a heterogeneous region of the sample reveals differences in local properties of the exciton as shown in Fig. \ref{fig1}(b-d). This additionally allows characterization of sample homogeneity to determine the optimal location of the pump and probe beams and can be used to correlate the inhomogeneous linewidth of the excitonic response extracted from 2DES with the statistics of the mapped linear reflectivity spectra (Fig. S3).      

\begin{figure}[ht]
\centering
\includegraphics[width=1 \linewidth]{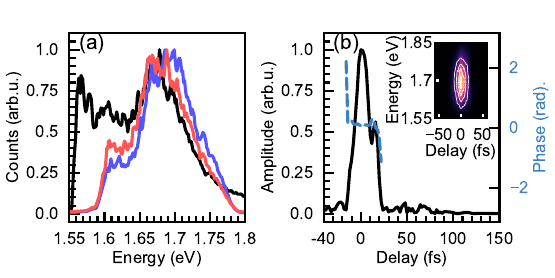}
\vspace{-10pt} 
\caption{Pulse characteristics at the sample position after going through the transmission objective. (a) fiber output spectrum (black, filtered < 800 nm) and shaped pump (blue) and probe (red) spectra. (b) Retrieved electric field amplitude and phase for the pump pulse at the sample position. Inset: tg-FROG trace of the pulse.  Retrieval algorithm: COPRA \cite{Geib2019} }
\label{fig2}
\end{figure}
Having recorded the linear properties of the sample, we move forward with the 2DES experiments. Here, the fiber was pumped with the laser at 1.55 eV. A spectrum of the fiber output after the 1.55 eV shortpass filter is shown in black in Fig. \ref{fig2}(a). This spectrum is subsequently shaped using pulse shapers in order to obtain a Gaussian-like spectrum centered around 1.667 eV, matching the center position of the A exciton (blue and red curves). An exemplary tg-FROG trace of the compressed beams at the sample position is shown as an inset of Fig. \ref{fig2}(b), while the retrieved field is shown in the main panel. A pulse duration of around 20 fs for both pump and probe is achieved. At the central energy of 1.667 eV, about 2900 fs$^2$ of chirp accumulated in the apochromat objective had to be compensated. The temporal profile of the pulse is exempt from tails or pedestals, which is made possible by compensating for phase distortions beyond second order with the pulse shaper.  
\begin{figure}[ht]
\centering
\includegraphics[width=1\linewidth]{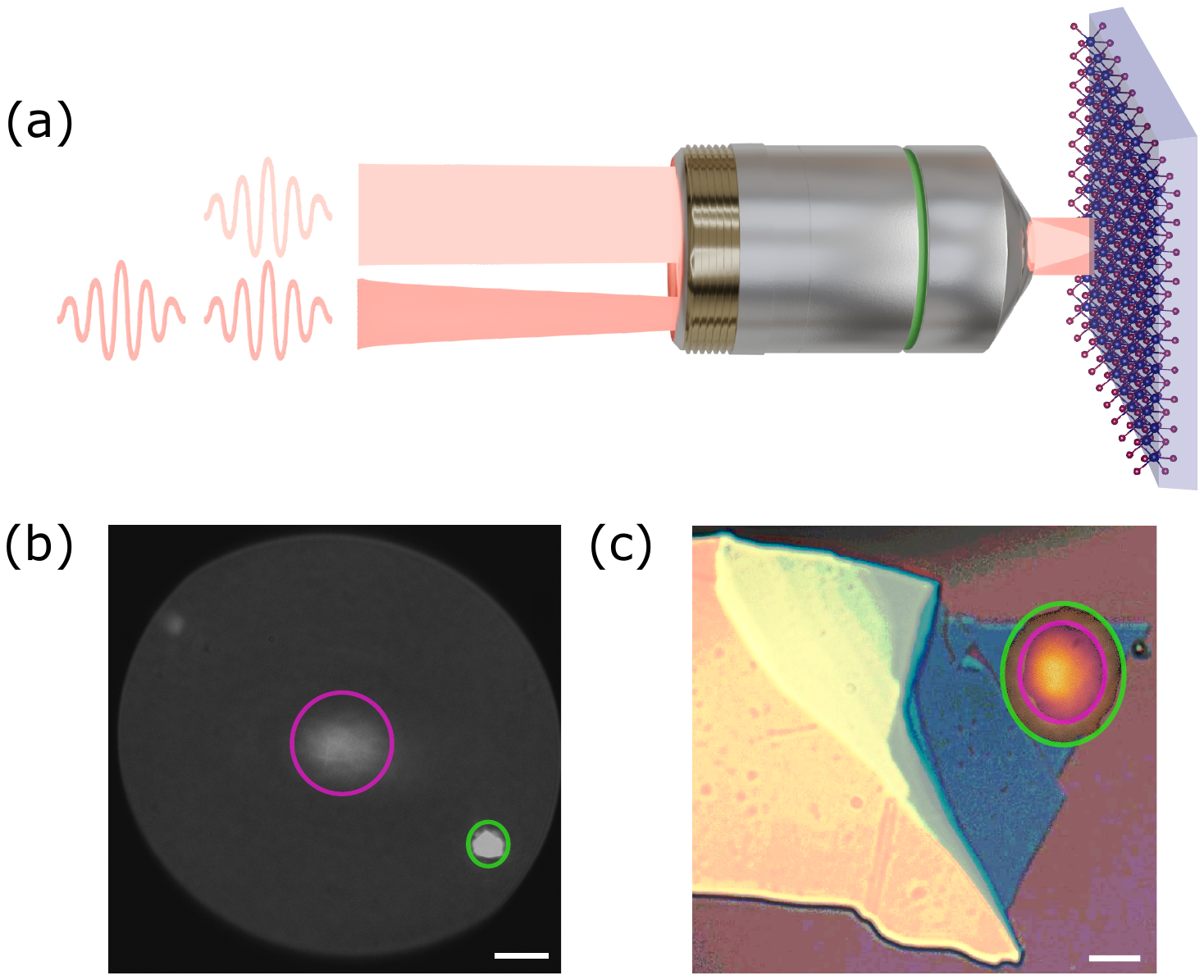}
\caption{Beam coupling into the microscope objective and principle of wavevector control. (a) Detailed schematics. (b) Image of the back focal plane (scale bar = 0.1 $\frac{k}{k_0}$) taken with CCD 2, with pump and probe beams circled in green and pink, respectively. (c) Corresponding real space image (scale bar = 10 $\mu m$), taken with CCD 1. Pump and probe beam spots are circled with matching colors to (b).}
\label{fig3}
\end{figure}

For the 2D experiment, the coherence time $t_1$ was scanned from 0 to 150 fs in steps of 1 fs. The population time was sampled
from -200 fs to 800 fs with 69 points in a partial rotating frame with a frequency of around 0.20 PHz. Each population time is averaged over 1000 independent shots.
 The experiment was carried out with pump energies of 500 nJ/pulse and a probe energy of 133 pJ/pulse. We use L2 and L3, as indicated in Fig. \ref{fig1}(a), to independently adjust the pump and probe spot sizes onto the back focal plane of the objective  (imaged using CCD 2). Having the possibility to independently tune pump and probe sizes can come in handy, for example, to rule out any diffusion effects arising from the pump. The principle is illustrated in more detail in Fig. \ref{fig3}(a). Here, the pump size was adjusted on the objective back focal plane using L3 to obtain a full-width at half-maximum spot size on the sample of around 14.9 $\mu m$ ($1/e^2$ = 25.3 $\mu m$), see Fig. \ref{fig3}(c). Its position on the back focal plane is shown in Fig. \ref{fig3}(b) as the green circle. Similarly, we use L2 to adjust the spot size of the probe. Here we employ a probe size with a full-width at half-maximum of 10.4 $\mu m$ ($1/e^2$ = 17.7 $\mu m$) (Fig. S2). Its position on the back focal plane of the objective is shown as the pink circle. In addition to controlling pump and probe sizes on the sample, this technique also permits working with pump and probe beams of the same color and polarization. While this way of coupling to the objective rules out diffraction-limited imaging, it still enables us to probe samples with sizes of 10-20 $\mu$m. Finally, imaging of the back focal plane enables us to precisely determine the wavevectors of the incident beams on the sample.

The use of an apochromat transmission-type objective enables access to any wavevector of light within the NA, in contrast to reflective-type objectives, where a significant part of the wavevectors are inaccessible due to their design. We anticipate that this feature will prove powerful to investigate light-momentum dependent 2D spectra in polaritonic systems \cite{Timmer2026, Li2022}. More generally, exciton lifetimes and optical properties of 2D materials were shown to depend on substrate or encapsulation layer thickness due to the Purcell effect \cite{Fang2019, Trofimov2025}. Having control over the angle of the light beams therefore, enables more systematic measurements of dynamical properties in 2D materials.

\begin{figure}[ht]
\centering
\includegraphics[width=0.7\linewidth]{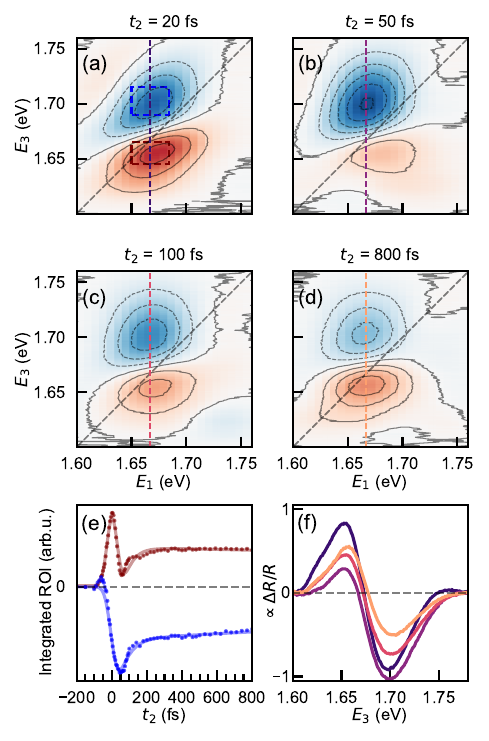}
\caption{Overview of 2D spectroscopy signals on monolayer WSe$_2$ on Si/SiO$_2$ (a-d) Exemplary 2D spectra at population times, $t_2$ = 20 fs, 50 fs, 100 fs and 800 fs, respectively. (e) Integrated temporal dynamics of spectral features shown by rectangles in (a), matching colors. (f) Pseudo transient reflectivity spectra at $E_1$ = 1.667 eV for different $t_2$.}
\label{fig4}
\end{figure}

Exemplary room temperature 2D spectra of the A exciton of WSe$_2$ are shown in Fig. \ref{fig4}(a-d), for selected population times. The spectra feature a derivative-like shape that is related to the reflection geometry. A cut at E$_1$ = 1.667 eV  (vertical dashed lines), corresponding to the center energy of the resonance, shows pseudo transient-resolved reflectivity spectra in Fig. \ref{fig4}(f). To assess the signal quality in more detail, we select a region of interest (ROI) along the blue and red features of the spectrum, shown as the dark blue and red rectangles in Fig. \ref{fig4}(a). The time-resolved traces obtained by integrating over the 2D signals within these ROIs are shown in Fig. \ref{fig4}(e). The trace in blue is fitted using a bi-exponential function convolved with a Gaussian to account for the finite instrument response function of the instrument. We recover a fast time constant of 60.0 $\pm$ 10.0 fs, and a slower time constant of 5.00 $\pm$ 0.75 ps. The fast time constant is comparable with previously measured ultrafast exciton dynamics in WSe$_2$ \cite{Mathias2013}.  Overall, the data in Fig. \ref{fig4} demonstrate the capabilities of our instruments to retrieve ultrafast dynamics on monolayer 2D materials.

In conclusion, we demonstrated the integration of confocal back-focal plane microscopy and 2D spectroscopy. Our setup retains the simplicity and flexibility of the pump-probe geometry,  can operate in both reflection and transmission geometries, and enables control over pump and probe wavevectors, of particular relevance for probing exciton-polaritons. The access to samples with sizes of 10-20 $\mu$m opens the door to several 2D materials of contemporary interest.

\section{Funding} We acknowledge funding from the DFG, under grant INST 130/1289-1 FUGG (Gross Geräte Antrag). The authors also acknowledge the DFG within Transregio TRR 227 Ultrafast Spin Dynamics (Project B11, Project-ID 328545488), within CRC 1772 on Heterostructures of Molecules and Two-Dimensional Materials (Project A02, Project ID No. 555467911), and within FOR 5750 (Project P4). This work was supported by the SupraFAB Research Center for sample production.

\section{Acknowledgment} We acknowledge Prof. Paul Fumagalli for providing the piezoelectric stages used for mapping linear reflectivity.

\section{Disclosures} The authors declare no conflicts of interest.

section{Data Availability Statement} The data shown in this paper will be available on Zenodo upon acceptance.

\section*{References}
\printbibliography[heading=none]

@article{Timmer2024,
  title = {Ultrafast Coherent Exciton Couplings and Many-Body Interactions in Monolayer WS2},
  volume = {24},
  ISSN = {1530-6992},
  url = {http://dx.doi.org/10.1021/acs.nanolett.4c01991},
  DOI = {10.1021/acs.nanolett.4c01991},
  number = {26},
  journal = {Nano Letters},
  publisher = {American Chemical Society (ACS)},
  author = {Timmer,  Daniel and Gittinger,  Moritz and Quenzel,  Thomas and Cadore,  Alisson R. and Rosa,  Barbara L. T. and Li,  Wenshan and Soavi,  Giancarlo and L\"{u}nemann,  Daniel C. and Stephan,  Sven and Silies,  Martin and Schulz,  Tommy and Steinhoff,  Alexander and Jahnke,  Frank and Cerullo,  Giulio and Ferrari,  Andrea C. and De Sio,  Antonietta and Lienau,  Christoph},
  year = {2024},
  month = jun,
  pages = {8117–8125}
}

@article{Hao2016,
  title = {Direct measurement of exciton valley coherence in monolayer WSe2},
  volume = {12},
  ISSN = {1745-2481},
  url = {http://dx.doi.org/10.1038/nphys3674},
  DOI = {10.1038/nphys3674},
  number = {7},
  journal = {Nature Physics},
  publisher = {Springer Science and Business Media LLC},
  author = {Hao,  Kai and Moody,  Galan and Wu,  Fengcheng and Dass,  Chandriker Kavir and Xu,  Lixiang and Chen,  Chang-Hsiao and Sun,  Liuyang and Li,  Ming-Yang and Li,  Lain-Jong and MacDonald,  Allan H. and Li,  Xiaoqin},
  year = {2016},
  month = feb,
  pages = {677–682}
}

@article{Guo2018,
  title = {Exchange-driven intravalley mixing of excitons in monolayer transition metal dichalcogenides},
  volume = {15},
  ISSN = {1745-2481},
  url = {http://dx.doi.org/10.1038/s41567-018-0362-y},
  DOI = {10.1038/s41567-018-0362-y},
  number = {3},
  journal = {Nature Physics},
  publisher = {Springer Science and Business Media LLC},
  author = {Guo,  Liang and Wu,  Meng and Cao,  Ting and Monahan,  Daniele M. and Lee,  Yi-Hsien and Louie,  Steven G. and Fleming,  Graham R.},
  year = {2018},
  month = dec,
  pages = {228–232}
}

@article{Lloyd2021,
  title = {Sub-10 fs Intervalley Exciton Coupling in Monolayer MoS2 Revealed by Helicity-Resolved Two-Dimensional Electronic Spectroscopy},
  volume = {15},
  ISSN = {1936-086X},
  url = {http://dx.doi.org/10.1021/acsnano.1c02381},
  DOI = {10.1021/acsnano.1c02381},
  number = {6},
  journal = {ACS Nano},
  publisher = {American Chemical Society (ACS)},
  author = {Lloyd,  Lawson T. and Wood,  Ryan E. and Mujid,  Fauzia and Sohoni,  Siddhartha and Ji,  Karen L. and Ting,  Po-Chieh and Higgins,  Jacob S. and Park,  Jiwoong and Engel,  Gregory S.},
  year = {2021},
  month = jun,
  pages = {10253–10263}
}

@article{Moody2015,
  title = {Intrinsic homogeneous linewidth and broadening mechanisms of excitons in monolayer transition metal dichalcogenides},
  volume = {6},
  ISSN = {2041-1723},
  url = {http://dx.doi.org/10.1038/ncomms9315},
  DOI = {10.1038/ncomms9315},
  number = {1},
  journal = {Nature Communications},
  publisher = {Springer Science and Business Media LLC},
  author = {Moody,  Galan and Kavir Dass,  Chandriker and Hao,  Kai and Chen,  Chang-Hsiao and Li,  Lain-Jong and Singh,  Akshay and Tran,  Kha and Clark,  Genevieve and Xu,  Xiaodong and Bergh\"{a}user,  Gunnar and Malic,  Ermin and Knorr,  Andreas and Li,  Xiaoqin},
  year = {2015},
  month = sep 
}

@article{Purz2021,
  title = {Coherent exciton-exciton interactions and exciton dynamics in a MoSe2/WSe2 heterostructure},
  volume = {104},
  ISSN = {2469-9969},
  url = {http://dx.doi.org/10.1103/PhysRevB.104.L241302},
  DOI = {10.1103/physrevb.104.l241302},
  number = {24},
  journal = {Physical Review B},
  publisher = {American Physical Society (APS)},
  author = {Purz,  Torben L. and Martin,  Eric W. and Rivera,  Pasqual and Holtzmann,  William G. and Xu,  Xiaodong and Cundiff,  Steven T.},
  year = {2021},
  month = dec 
}

@article{Purz2022,
  title = {Imaging dynamic exciton interactions and coupling in transition metal dichalcogenides},
  volume = {156},
  ISSN = {1089-7690},
  url = {http://dx.doi.org/10.1063/5.0087544},
  DOI = {10.1063/5.0087544},
  number = {21},
  journal = {The Journal of Chemical Physics},
  publisher = {AIP Publishing},
  author = {Purz,  Torben L. and Martin,  Eric W. and Holtzmann,  William G. and Rivera,  Pasqual and Alfrey,  Adam and Bates,  Kelsey M. and Deng,  Hui and Xu,  Xiaodong and Cundiff,  Steven T.},
  year = {2022},
  month = jun 
}

@article{Policht2021,
  title = {Dissecting Interlayer Hole and Electron Transfer in Transition Metal Dichalcogenide Heterostructures via Two-Dimensional Electronic Spectroscopy},
  volume = {21},
  ISSN = {1530-6992},
  url = {http://dx.doi.org/10.1021/acs.nanolett.1c01098},
  DOI = {10.1021/acs.nanolett.1c01098},
  number = {11},
  journal = {Nano Letters},
  publisher = {American Chemical Society (ACS)},
  author = {Policht,  Veronica R. and Russo,  Mattia and Liu,  Fang and Trovatello,  Chiara and Maiuri,  Margherita and Bai,  Yusong and Zhu,  Xiaoyang and Dal Conte,  Stefano and Cerullo,  Giulio},
  year = {2021},
  month = may,
  pages = {4738–4743}
}

@article{Li2022,
  title = {Hybridized Exciton-Photon-Phonon States in a Transition Metal Dichalcogenide van der Waals Heterostructure Microcavity},
  volume = {128},
  ISSN = {1079-7114},
  url = {http://dx.doi.org/10.1103/PhysRevLett.128.087401},
  DOI = {10.1103/physrevlett.128.087401},
  number = {8},
  journal = {Physical Review Letters},
  publisher = {American Physical Society (APS)},
  author = {Li,  Donghai and Shan,  Hangyong and Rupprecht,  Christoph and Knopf,  Heiko and Watanabe,  Kenji and Taniguchi,  Takashi and Qin,  Ying and Tongay,  Sefaattin and Nuß,  Matthias and Schr\"{o}der,  Sven and Eilenberger,  Falk and H\"{o}fling,  Sven and Schneider,  Christian and Brixner,  Tobias},
  year = {2022},
  month = feb 
}

@article{Baiz2014,
  title = {Ultrafast 2D IR microscopy},
  volume = {22},
  ISSN = {1094-4087},
  url = {http://dx.doi.org/10.1364/OE.22.018724},
  DOI = {10.1364/oe.22.018724},
  number = {15},
  journal = {Optics Express},
  publisher = {Optica Publishing Group},
  author = {Baiz,  Carlos R. and Schach,  Denise and Tokmakoff,  Andrei},
  year = {2014},
  month = jul,
  pages = {18724}
}

@article{Goetz2018,
  title = {Coherent two-dimensional fluorescence micro-spectroscopy},
  volume = {26},
  ISSN = {1094-4087},
  url = {http://dx.doi.org/10.1364/OE.26.003915},
  DOI = {10.1364/oe.26.003915},
  number = {4},
  journal = {Optics Express},
  publisher = {Optica Publishing Group},
  author = {Goetz,  Sebastian and Li,  Donghai and Kolb,  Verena and Pflaum,  Jens and Brixner,  Tobias},
  year = {2018},
  month = feb,
  pages = {3915}
}

@article{Steves2019,
  title = {Correlated spatially resolved two-dimensional electronic and linear absorption spectroscopy},
  volume = {44},
  ISSN = {1539-4794},
  url = {http://dx.doi.org/10.1364/OL.44.002117},
  DOI = {10.1364/ol.44.002117},
  number = {8},
  journal = {Optics Letters},
  publisher = {Optica Publishing Group},
  author = {Steves,  Megan A. and Zheng,  Hongjun and Knappenberger,  Kenneth L.},
  year = {2019},
  month = apr,
  pages = {2117}
}

@article{Gross2023,
  title = {Progress and Prospects in Optical Ultrafast Microscopy in the Visible Spectral Region: Transient Absorption and Two-Dimensional Microscopy},
  volume = {127},
  ISSN = {1932-7455},
  url = {http://dx.doi.org/10.1021/acs.jpcc.3c02091},
  DOI = {10.1021/acs.jpcc.3c02091},
  number = {30},
  journal = {The Journal of Physical Chemistry C},
  publisher = {American Chemical Society (ACS)},
  author = {Gross,  Niklas and Kuhs,  Christopher T. and Ostovar,  Behnaz and Chiang,  Wei-Yi and Wilson,  Kelly S. and Volek,  Tanner S. and Faitz,  Zachary M. and Carlin,  Claire C. and Dionne,  Jennifer A. and Zanni,  Martin T. and Gruebele,  Martin and Roberts,  Sean T. and Link,  Stephan and Landes,  Christy F.},
  year = {2023},
  month = jul,
  pages = {14557–14586}
}

@article{Ostrander2016,
  title = {Spatially Resolved Two-Dimensional Infrared Spectroscopy via Wide-Field Microscopy},
  volume = {3},
  ISSN = {2330-4022},
  url = {http://dx.doi.org/10.1021/acsphotonics.6b00297},
  DOI = {10.1021/acsphotonics.6b00297},
  number = {7},
  journal = {ACS Photonics},
  publisher = {American Chemical Society (ACS)},
  author = {Ostrander,  Joshua S. and Serrano,  Arnaldo L. and Ghosh,  Ayanjeet and Zanni,  Martin T.},
  year = {2016},
  month = jun,
  pages = {1315–1323}
}

@article{Jones2019,
  title = {Multidimensional Spectroscopy on the Microscale: Development of a Multimodal Imaging System Incorporating 2D White-Light Spectroscopy,  Broadband Transient Absorption,  and Atomic Force Microscopy},
  volume = {123},
  ISSN = {1520-5215},
  url = {http://dx.doi.org/10.1021/acs.jpca.9b09099},
  DOI = {10.1021/acs.jpca.9b09099},
  number = {50},
  journal = {The Journal of Physical Chemistry A},
  publisher = {American Chemical Society (ACS)},
  author = {Jones,  Andrew C. and Kearns,  Nicholas M. and Bohlmann Kunz,  Miriam and Flach,  Jessica T. and Zanni,  Martin T.},
  year = {2019},
  month = nov,
  pages = {10824–10836}
}

@article{Tiwari2018,
  title = {Spatially-resolved fluorescence-detected two-dimensional electronic spectroscopy probes varying excitonic structure in photosynthetic bacteria},
  volume = {9},
  ISSN = {2041-1723},
  url = {http://dx.doi.org/10.1038/s41467-018-06619-x},
  DOI = {10.1038/s41467-018-06619-x},
  number = {1},
  journal = {Nature Communications},
  publisher = {Springer Science and Business Media LLC},
  author = {Tiwari,  Vivek and Matutes,  Yassel Acosta and Gardiner,  Alastair T. and Jansen,  Thomas L. C. and Cogdell,  Richard J. and Ogilvie,  Jennifer P.},
  year = {2018},
  month = oct 
}

@article{Timmer2026,
  title = {Ultrafast transition from coherent to incoherent polariton nonlinearities in a hybrid 1L-WS2/plasmon structure},
  ISSN = {1748-3395},
  url = {http://dx.doi.org/10.1038/s41565-025-02054-4},
  DOI = {10.1038/s41565-025-02054-4},
  journal = {Nature Nanotechnology},
  publisher = {Springer Science and Business Media LLC},
  author = {Timmer,  Daniel and Gittinger,  Moritz and Quenzel,  Thomas and Cadore,  Alisson R. and Rosa,  Barbara L. T. and Li,  Wenshan and Soavi,  Giancarlo and L\"{u}nemann,  Daniel C. and Stephan,  Sven and Greten,  Lara and Richter,  Marten and Knorr,  Andreas and De Sio,  Antonietta and Silies,  Martin and Cerullo,  Giulio and Ferrari,  Andrea C. and Lienau,  Christoph},
  year = {2026},
  month = jan 
}

@article{Pawowska2014,
  title = {Shaping and spatiotemporal characterization of sub-10-fs pulses focused by a high-NA objective},
  volume = {22},
  ISSN = {1094-4087},
  url = {http://dx.doi.org/10.1364/OE.22.031496},
  DOI = {10.1364/oe.22.031496},
  number = {25},
  journal = {Optics Express},
  publisher = {Optica Publishing Group},
  author = {Pawłowska,  Monika and Goetz,  Sebastian and Dreher,  Christian and Wurdack,  Matthias and Krauss,  Enno and Razinskas,  Gary and Geisler,  Peter and Hecht,  Bert and Brixner,  Tobias},
  year = {2014},
  month = dec,
  pages = {31496}
}

@article{Seiler2017,
  title = {Simple fiber-based solution for coherent multidimensional spectroscopy in the visible regime},
  volume = {42},
  ISSN = {1539-4794},
  url = {http://dx.doi.org/10.1364/OL.42.000643},
  DOI = {10.1364/ol.42.000643},
  number = {3},
  journal = {Optics Letters},
  publisher = {Optica Publishing Group},
  author = {Seiler,  Hélène and Palato,  Samuel and Schmidt,  Bruno E. and Kambhampati,  Patanjali},
  year = {2017},
  month = feb,
  pages = {643}
}

@article{Sonnichsen2021,
  title = {OPA-driven hollow-core fiber as a tunable,  broadband source for coherent multidimensional spectroscopy},
  volume = {29},
  ISSN = {1094-4087},
  url = {http://dx.doi.org/10.1364/OE.431988},
  DOI = {10.1364/oe.431988},
  number = {18},
  journal = {Optics Express},
  publisher = {Optica Publishing Group},
  author = {Sonnichsen,  Colin and Brosseau,  Patrick and Reid,  Cameron and Kambhampati,  Patanjali},
  year = {2021},
  month = aug,
  pages = {28352}
}

@article{Timmer2023,
  title = {Full visible range two-dimensional electronic spectroscopy with high time resolution},
  volume = {32},
  ISSN = {1094-4087},
  url = {http://dx.doi.org/10.1364/OE.511906},
  DOI = {10.1364/oe.511906},
  number = {1},
  journal = {Optics Express},
  publisher = {Optica Publishing Group},
  author = {Timmer,  Daniel and L\"{u}nemann,  Daniel C. and Riese,  Sebastian and Sio,  Antonietta De and Lienau,  Christoph},
  year = {2023},
  month = dec,
  pages = {835}
}

@article{Ma2016,
  title = {Broadband 7-fs diffractive-optic-based 2D electronic spectroscopy using hollow-core fiber compression},
  volume = {24},
  ISSN = {1094-4087},
  url = {http://dx.doi.org/10.1364/OE.24.020781},
  DOI = {10.1364/oe.24.020781},
  number = {18},
  journal = {Optics Express},
  publisher = {Optica Publishing Group},
  author = {Ma,  Xiaonan and Dostál,  Jakub and Brixner,  Tobias},
  year = {2016},
  month = aug,
  pages = {20781}
}

@article{AlHaddad2015,
  title = {Set-up for broadband Fourier-transform multidimensional electronic spectroscopy},
  volume = {40},
  ISSN = {1539-4794},
  url = {http://dx.doi.org/10.1364/OL.40.000312},
  DOI = {10.1364/ol.40.000312},
  number = {3},
  journal = {Optics Letters},
  publisher = {Optica Publishing Group},
  author = {Al Haddad,  A. and Chauvet,  A. and Ojeda,  J. and Arrell,  C. and van Mourik,  F. and Aub\"{o}ck,  G. and Chergui,  M.},
  year = {2015},
  month = jan,
  pages = {312}
}

@article{Verluise2000,
  title = {Amplitude and phase control of ultrashort pulses by use of an acousto-optic programmable dispersive filter: pulse compression and shaping},
  volume = {25},
  ISSN = {1539-4794},
  url = {http://dx.doi.org/10.1364/OL.25.000575},
  DOI = {10.1364/ol.25.000575},
  number = {8},
  journal = {Optics Letters},
  publisher = {Optica Publishing Group},
  author = {Verluise,  F. and Laude,  V. and Cheng,  Z. and Spielmann,  Ch. and Tournois,  P.},
  year = {2000},
  month = apr,
  pages = {575}
}

@article{Loriot2013,
  title = {Self-referenced characterization of femtosecond laser pulses by chirp scan},
  volume = {21},
  ISSN = {1094-4087},
  url = {http://dx.doi.org/10.1364/OE.21.024879},
  DOI = {10.1364/oe.21.024879},
  number = {21},
  journal = {Optics Express},
  publisher = {Optica Publishing Group},
  author = {Loriot,  Vincent and Gitzinger,  Gregory and Forget,  Nicolas},
  year = {2013},
  month = oct,
  pages = {24879}
}

@article{Trebino1997,
  title = {Measuring ultrashort laser pulses in the time-frequency domain using frequency-resolved optical gating},
  volume = {68},
  ISSN = {1089-7623},
  url = {http://dx.doi.org/10.1063/1.1148286},
  DOI = {10.1063/1.1148286},
  number = {9},
  journal = {Review of Scientific Instruments},
  publisher = {AIP Publishing},
  author = {Trebino,  Rick and DeLong,  Kenneth W. and Fittinghoff,  David N. and Sweetser,  John N. and Krumb\"{u}gel,  Marco A. and Richman,  Bruce A. and Kane,  Daniel J.},
  year = {1997},
  month = sept,
  pages = {3277–3295}
}

@article{Shim2009,
  title = {How to turn your pump–probe instrument into a multidimensional spectrometer: 2D IR and Vis spectroscopiesvia pulse shaping},
  volume = {11},
  ISSN = {1463-9084},
  url = {http://dx.doi.org/10.1039/B813817F},
  DOI = {10.1039/b813817f},
  number = {5},
  journal = {Phys. Chem. Chem. Phys.},
  publisher = {Royal Society of Chemistry (RSC)},
  author = {Shim,  Sang-Hee and Zanni,  Martin T.},
  year = {2009},
  pages = {748–761}
}

@article{Myers2008,
  title = {Two-color two-dimensional Fourier transform electronic spectroscopy with a pulse-shaper},
  volume = {16},
  ISSN = {1094-4087},
  url = {http://dx.doi.org/10.1364/oe.16.017420},
  DOI = {10.1364/oe.16.017420},
  number = {22},
  journal = {Optics Express},
  publisher = {Optica Publishing Group},
  author = {Myers,  Jeffrey A. and Lewis,  Kristin L. and Tekavec,  Patrick F. and Ogilvie,  Jennifer P.},
  year = {2008},
  month = oct,
  pages = {17420}
}

@misc{Azarm2026_arxiv,
Author = {Mohammadjavad Azarm and Rizwan Asif and Alessandra Milloch and Donna Datta and Ambrine Lanseur and Filippo Fabbri and Federica Bianco and Fabrizio Preda and Antonio Perri and Giulio Cerullo and Stefania Pagliara and Gabriele Ferrini and Claudio Giannetti},
Title = {Coherent multi-dimensional widefield microscopy},
Year = {2026},
Eprint = {arXiv:2603.23759},
}

@article{Fang2019,
  title = {Control of the Exciton Radiative Lifetime in van der Waals Heterostructures},
  volume = {123},
  ISSN = {1079-7114},
  url = {http://dx.doi.org/10.1103/PhysRevLett.123.067401},
  DOI = {10.1103/physrevlett.123.067401},
  number = {6},
  journal = {Physical Review Letters},
  publisher = {American Physical Society (APS)},
  author = {Fang,  H. H. and Han,  B. and Robert,  C. and Semina,  M. A. and Lagarde,  D. and Courtade,  E. and Taniguchi,  T. and Watanabe,  K. and Amand,  T. and Urbaszek,  B. and Glazov,  M. M. and Marie,  X.},
  year = {2019},
  month = Aug 
}

@article{Trofimov2025,
  title = {Directed light emission from monolayers on 2D materials via optical interferences},
  volume = {163},
  ISSN = {1089-7690},
  url = {http://dx.doi.org/10.1063/5.0279864},
  DOI = {10.1063/5.0279864},
  number = {8},
  journal = {The Journal of Chemical Physics},
  publisher = {AIP Publishing},
  author = {Trofimov,  P. and Juergensen,  S. and Dewambrechies Fernández,  A. and Bolotin,  K. I. and Reich,  S. and Seiler,  H.},
  year = {2025},
  month = Aug 
}

@article{Geib2019,
  title = {Common pulse retrieval algorithm: a fast and universal method to retrieve ultrashort pulses},
  volume = {6},
  ISSN = {495-505},
  url = {https://doi.org/10.1364/OPTICA.6.000495},
  DOI = {10.1364/OPTICA.6.000495},
  number = {4},
  journal = {Optica},
  publisher = {Optica Publishing Group},
  author = {Geib,  Nils C. and Zilk,  M. and Pertach,  T. and Eilenberger,  F.},
  year = {2019},
  month = April 
}

@article{Mathias2013,
  title = {Ultrafast dynamics of bright and dark excitons in monolayer WSe2 and heterobilayer WSe2/MoS2},
  volume = {10},
  ISSN = {2053-1583},
  url = {https://iopscience.iop.org/article/10.1088/2053-1583/ace067/meta},
  DOI = {10.1088/2053-1583/ace067},
  number = {3},
  journal = {2D Materials},
  publisher = { IOP Publishing Ltd.},
author = {Bange, Jan Philipp and Werner, Paul and Schmitt, David and Bennecke, Wiebke and Meneghini, Giuseppe and AlMutairi, AbdulAziz and Merboldt, Marco and Watanabe, Kenji and Taniguchi, Takashi and Steil, Sabine and Steil, Daniel and Weitz, R Thomas and Hofmann, Stephan and Jansen, G S Matthijs and Brem, Samuel and Malic, Ermin and Reutzel, Marcel and Mathias, Stefan},
  year = {2023},
  month = July 
}

@article{Novoselov2016,
  title = {2D materials and van der Waals heterostructures},
  volume = {353},
  ISSN = {1095-9203},
  url = {http://dx.doi.org/10.1126/science.aac9439},
  DOI = {10.1126/science.aac9439},
  number = {6298},
  journal = {Science},
  publisher = {American Association for the Advancement of Science (AAAS)},
  author = {Novoselov,  K. S. and Mishchenko,  A. and Carvalho,  A. and Castro Neto,  A. H.},
  year = {2016},
  month = July 
}

@article{Xu2023,
  title = {Ultrafast imaging of polariton propagation and interactions},
  volume = {14},
  ISSN = {2041-1723},
  url = {http://dx.doi.org/10.1038/s41467-023-39550-x},
  DOI = {10.1038/s41467-023-39550-x},
  number = {1},
  journal = {Nature Communications},
  publisher = {Springer Science and Business Media LLC},
  author = {Xu,  Ding and Mandal,  Arkajit and Baxter,  James M. and Cheng,  Shan-Wen and Lee,  Inki and Su,  Haowen and Liu,  Song and Reichman,  David R. and Delor,  Milan},
  year = {2023},
  month = June 
}

@book{zanni,
    author = {Hamm, Peter and Zanni , Martin},
    title = {Concepts and Methods of 2D Infrared Spectroscopy},
    publisher = {Cambridge University Press},
    year = {2012},
    isbn = {9780511675935},
    url = {https://doi.org/10.1017/CBO9780511675935}
}

\end{document}